\documentstyle[twocolumn,aps]{revtex}
\input{epsf.tex}

\def\be{\begin{equation}}
\def\ee{\end{equation}}
\def\ba{\begin{array}}
\def\ea{\end{array}}
\def\bea{\begin{eqnarray}}
\def\eea{\end{eqnarray}}
\begin{document}
\draft
\title{\bf A new semiempirical formula for exotic cluster decays of nuclei}
\author{{M. Balasubramaniam, S. Kumarasamy and N. Arunachalam}\\
{Department of Physics, Manonmaniam Sundaranar University, Tirunelveli-627012, India.}\\
}
\author{{Raj K. Gupta}\\
{Department of Physics, Panjab University, Chandigarh-160014, India.}\\
}
\date{\today}
\maketitle

\begin{abstract}
A new semiempirical formula, with only three parameters, is proposed for cluster decay 
half-lives. The parameters of the formula are obtained by making a least squares fit to 
the available experimental data. The calculated half-lives are compared with an earlier
proposed model-independent scaling law. Also, the calculated results of this formula are 
compared with the recent results of the preformed cluster model for $^{12}$C and $^{14}$C 
emissions from different deformed and superdeformed Nd and Gd parents. The results are 
in good agreement with experiments as well as other models.
\end{abstract}

\section{Introduction}
In a radioactive decay series, the end product is reached not only via the emission of
$\alpha$ and $\beta$ particles but also directly via heavy nuclei emission, the clusters,
like Carbon, Oxygen, Florine, Neon, Magnesium and Silicon. Such a process is known as
cluster decay, first proposed theoretically in 1980 by S\v andulescu, Poenaru and Greiner
\cite{san80}, before its experimental realization in 1984 by Rose and Jones \cite{ros84}.
In this decay process a parent nucleus (A,Z), with mass number A and charge number Z,
breaks into two fragments, \textit{viz.,} the emitted cluster (A$_2$,Z$_2$) and the
associated daughter (A$_1$,Z$_1$), where A=A$_1$+A$_2$, Z=Z$_1$+Z$_2$. The light fragment
(A$_2$,Z$_2$) is a cluster, heavier than the $\alpha$-particle but lighter than the
lightest fission fragment observed so far.

The detection of cluster-decay was hindered mainly due to large pile-up of $\alpha$
particles, but with improved facilities these difficulties are removed and a large number
of cluster-decays are observed from different radioactive nuclei. In the years that
followed, the cluster decay process has been studied extensively using different theoretical
models with different realistic nuclear interaction potentials. In general, two kinds
of models are used for explaining the observed and/ or for predicting new decay modes.
In one kind of these models, the $\alpha$-particle as well as the heavy cluster(s) were
assumed to be pre-born in a parent nucleus, before they could penetrate the barrier with
the available Q-value. These models are called the preformed cluster models
\cite{rkg88,mal89,kum97,ble87,ble88}. In such a model, the clusters of different sizes
(mass and/ or charge numbers) are considered to be preformed in the parent nucleus, with
different probabilities. The Gamow-like barrier penetration is also taken into account
in these models. In the other kind of models, only the Gamow's idea of barrier penetration
is used, without considering the cluster(s) being or not being preformed in the parent
nucleus. In other words, in this kind of models, called the unified fission models
\cite{san80,shi85,pik86,sha88,buc89,san92,royer98}, the cluster radioactivity is considered as
a simple barrier penetration phenomenon, in between the $\alpha$-decay and the spontaneous
fission. In this paper, we attempt to give a model-independent, semiempirical formula for
studying this above mentioned process of exotic cluster decay.

Geiger and Nuttal \cite{geiger11} were the first who proposed a semiempirical law
connecting the $\alpha$-decay half-life and its Q-value. Now, a large number of
$\alpha$-decays are observed from medium mass to superheavy nuclei, and several attempts
have been made to give a universal formula \cite{hor94,bro92,roy00}. These formulae vary
among themselves mainly in the number of parameters. One such scaling law, proposed
recently by Horoi {\it et al} \cite{hor94}, accounts for both the $\alpha$ and
cluster decays, and is given as,
\begin{equation}
\log
T_{1/2}=(a_1\mu^x+b_1)\left[\frac{(Z_1Z_2)^y}{\sqrt{Q}}-7\right]+(a_2\mu^x+b_2).
\label{eq:1}
\end{equation}
This scaling law involves six parameters $a_1=9.1$, $b_1=-10.2$, $a_2=7.39$, 
$b_2=-23.2$, $x=0.416$ and $y=0.613$, which were obtained by fitting 119 
$\alpha$-decays and 11 cluster-decays from various even-even parents. In the following, 
we propose a new semiempirical formula based on only three parameters that are
least squares fitted to the cluster-decay data alone. So far, no attempt has been made 
to fit the $\alpha$-decay data.

The evolution of our proposed formula is presented in Section II, followed by a brief
description of the preformed cluster model (PCM) of Gupta and collaborators
\cite{rkg88,mal89,kum97} in Section III. The PCM is recently used by two of us and
collaborators \cite{gupta03} to calculate the cluster-decay half-lives of some deformed and
superdeformed Gd and Nd parents. Some of these results of Ref. \cite{gupta03} are
used here for comparisons with the calculations based on the proposed new semiempirical
formula. The results of our calculation are discussed in Section IV, and the summary and 
conclusions are presented in Section V.

\section{The new semiempirical formula}
We base the new formula for cluster decay half-lives on the following three simple
experimental facts: \\
\begin{description}
\item{(i)} It is known from experiments that the cluster decay half-lives increase
with the size (mass and/or charge) of the clusters. Hence, the empirical formula should
contain terms showing direct dependence on the mass number and charge number of the cluster.
\item{(ii)} The same cluster is emitted by different parents and hence the formula should
contain dependence on the mass and charge asymmetries
\begin{equation}
\eta=\frac{A_1-A_2}{A}; \qquad \eta_z=\frac{Z_1-Z_2}{Z},
\label{eq:2}
\end{equation}
respectively.
\item{(iii)} Since the $\alpha$-decay and cluster-decay are physically similar processes,
the Q-dependence is taken to be the same as in Geiger-Nuttal law for $\alpha$-decay, i.e.,
$\log T_{1/2}\propto Q^{-1/2}$.
\end{description}
Combining the above three results, we get
\begin{equation}
\log T_{1/2}^{AZ}=\frac{a A_2 \eta+b Z_2 \eta_z}{\sqrt{Q}}+c,
\label{eq:3}
\end{equation}
where the constants $a=10.603$, $b=78.027$ and $c=-80.669$ are obtained by making a least 
squares fit of the available experimental half-lives for exotic cluster decays alone, with 
an rms deviation $d_{rms}$=0.89(s), defined as
$d_{rms}=\sum_{i=1}^n\left[\frac{y_i-f(x_i,a_j)}{\sigma_i}\right]^2$, with $f(x_i,a_j)$ 
denoting the function in Eq. (\ref{eq:3}), n the number of measurements and $y_i$ the 
experimentally observed values. The 
$\sigma_i^2$=$\frac{1}{n}\sum_{i=1}^n(y_i-\bar{y})^2$, 
the variance, gives the standard deviation $\sigma_i$, with 
$\bar{y}=\frac{1}{n}\sum_{i=1}^n y_i$ 
giving the arithmetic average of the expermentally measured quantities. We refer to this 
formula as the \textit{AZ-formula} (\textit{AZF}). Note that the dependence on both 
$\eta$ and $\eta_Z$ must be included in the formula (\ref{eq:3}) since these are 
separately measurable quantities, though the coupling between them is known to be weak
\cite{maruhn74,gupta75,gupta99}, as is also evident when we consider the $\eta$ and 
$\eta_Z$ dependence separately. These special cases of AZ-formula, i.e., $b$=0 or $a$=0 
in (\ref{eq:3}), respectively, are expected to give reasonably good results, though
poorer than the AZF results. These truncated expressions are referred to as the 
\textit{AF} and \textit{ZF} formulae in the following, whose constants are also obtained 
directly by the least squares fit to data. These are: $a=30.568$ and $c=-51.348$ for AF
and  $b=112.197$ and $c=-89.025$ for ZF, with rms deviations $d_{rms}$=1.652 and 1.112 (s), 
respectively. Apparently, the rms deviations for truncated expressions are larger, and 
hence the fits are poorer, compared to the total expression (\ref{eq:3}). This will also be
evident when we compare the results of these expressions with experimental data in 
Section IV.

\section{The preformed cluster model}
In the prefomed cluster model (PCM) of Gupta and collaborators \cite{rkg88,mal89,kum97},
the decay half-life is given in terms of three factors, the P$_0$, P and $\nu$, as
\begin{equation}
\log T_{1/2}=ln 2/P_{0}P{\nu},
\label{eq:4}
\end{equation}
where $P_{0}$ is the preformation probability of the two fragments (the cluster and
daughter nuclei) in their respective ground states, P the probability to tunnel the 
confining nuclear interaction barrier and $\nu$ as an assault frequency.

For calculating $P_0$ and P, Gupta {\it et al.} introduced the coupled motion in dynamical
collective coordinates of mass asymmetry $\eta$ and relative separation R via a
stationary Schr\"{o}dinger equation
\begin{equation}
H(\eta,R){\psi(\eta,R)}=E{\psi(\eta,R)},
\label{eq:5}
\end{equation}
with the potential in it defined by the sum of experimental binding energies \cite{audi95},
the Coulomb and the proximity \cite{blocki77} potentials, as
\begin{equation}
V(\eta,R)=\sum_{i=1}^2  B_i(A_i,Z_i)+\frac{Z_1Z_2e^2}{R}+V_P.
\label{eq:6}
\end{equation}
Here the charges Z$_i$ are fixed by minimizing the potential (without V$_P$) in the charge
asymmetry co-ordinate $\eta_{z}$. Equation (\ref{eq:5}) is solved in the decoupled
approximation, which gives
\begin{equation}
P_0 \propto {|\psi(\eta)|^2}
\label{eq:7}
\end{equation}
\begin{equation}
P \propto {|\psi(R)|^2}.
\label{eq:8}
\end{equation}
Only the ground state solution is relevant for the cluster decay to occur in the ground
state of the daughter nucleus. Then, for $\eta$-motion, the properly normalized fractional
preformation probability for, say, cluster A$_2$ at a fixed R (=R$_a$=C$_t$=C$_1$+C$_2$,
C$_i$ being the S\"ussmann central radii) is
\begin{equation}
P_0(A_2)={|\psi(\eta)|^2}{\sqrt{B_{\eta\eta}}}\frac{2}{A}.
\label{eq:9}
\end{equation}
For R-motion, the potential V(R) is obtained from Eq. (\ref{eq:6}) for fixed $\eta$ and,
instead of solving the corresponding radial Schr\"{o}dinger equation, the WKB approximation
is used for calculating the penetrability P. Finally, the assault frequency $\nu$ in PCM
is defined by considering that the total kinetic energy, shared between the two fragments,
is the positive Q-value:
\begin{equation}
\nu=v/R_0=\frac{\sqrt{2Q/mA_2}}{R_0}.
\label{eq:10}
\end{equation}
Here $R_0$ is the spherical radius of the parent nucleus and $m$, the nucleon mass.\\

\section{Results and discussion}
Figure 1 and Table I give our calculated logarithms of decay half-lives by using the
\textit{AZ-formula} (\ref{eq:3}) for different clusters emitted from various radioactive
parents, compared with the experimental data. In Fig. 1, we have also plotted the results of
our calucations for \textit{AF} ($b=0$) and \textit{ZF} ($a=0$) versions of the
\textit{AZ-formula}. It is evident that the AZF fit to the data is better than the AF and ZF
fits, as expected from our discussion above in Section II.

In Table I, we have also added the experimental data on Q-values and the results of another
calculation using Eq. (\ref{eq:1}) of Horio et al. \cite{hor94}. The comparison between the
experiments and formulae (\ref{eq:1}) and (\ref{eq:3}) are also displayed in Fig. 2 for the
illustrative cases of $^{14}C$ and $^{24}Ne$ cluster decays (respectively, the upper and lower
panel). Apparently, our semiempirical \textit{AZ-formula} is much closer to experiments, as
compared to the other formula due to Horio et al. \cite{hor94}.

Finally, in order to compare the results of our semiempirical formula with the results of 
a model-dependent theoretical cluster decay calculation, we use a recent calculation 
\cite{gupta03} of two of us (MB and RKG) and collaborators based on PCM for the emission of 
$^{12}C$ and $^{14}C$ clusters from $^{133-137}_{60}$Nd and $^{144-158}_{64}$Gd parents. 
It may be mentioned here that these cluster-decay calculations are of interest only for 
nuclear structure information since their experimental observation may not be feasible in 
the near future \cite{gupta03}. Figs. 3(a) to (d) show the results of the PCM calculations, 
compared with those from the semiempirical \textit{AZ-formula} (\ref{eq:3}) and the scaling 
law (\ref{eq:1}) of Horoi {\it et al.} \cite{hor94}. We notice that, in general, the 
predictions of our semiempirical law lie higher than those of the PCM and Horoi {\it et al}. 
This is more so for Nd parents than for Gd parents where the predictions of the three 
calculations are nearly similar. The interesting point is that the three calculations predict 
an exactly the same structure for $logT_{1/2}$ vs. A, the parent mass number. For quantitative
comparisons, it may be reminded here that the constants a, b and c of our semiemprical 
formula are obtained by fitting the data from trans-actinide region, which may or may not 
be good for this trans-tin region.

\section{Summary}
In this paper, we have proposed a model-independent three parameter formula for calculating
the half-lives of cluster decays of nuclei. The evolution of the formula is based on three
simple experimental observations about the characteristics of exotic cluster decays. The
inputs of the formula are simply the mass and charge numbers of the parent and cluster, along
with the Q-value of decay. The predictions of the proposed formula are comparable to another
model-independent scaling law as well as to the well known preformed cluster model. The
resulting good comparisons suggest that this formula could be used to make predictions for
the guide of new experiments, similar to that of Poenaru {\it et al.} \cite{poe91}.

\begin{acknowledgments}
One of the authors (M.B.) acknowledges with thanks the partial
financial support by the Department of Science and Technology
(DST), vide Grant NO. SR/FTP/PSA-02/2002. Also, the support by DST
under the FIST programme vide letter No. SR/FST/PSI-005/2000 to
the Department of Physics, M.S. University, Tirunelveli, India, is
gratefully acknowledged.
\end{acknowledgments}

\vspace {1.0cm}

\par\noindent
{\bf Figure Captions}
\begin {description}
\item{Fig. 1} The logarithms of decay half-lives for different clusters emitted from various
radioactive parents, calculated by using {\it AZ-formula} (AZF) and compared with experimental
data. Also, the results of calculations for AF (b=0) and ZF (a=0) truncations of AZF are shown
for comparisons. The parents are labelled at the top X-axis and the corresponding clusters
emitted by these parents are labelled at the bottom X-axis.
\item{Fig. 2} The experimental data on logarithms of decay half-lives for the emission of
$^{14}C$ and $^{24}$Ne clusters from different radioactive parents, compared with the results
of calculations using {\it AZ-formula} (AZF) proposed here and the scaling law of Ref.
\cite{hor94}.
\item{Fig.3} The logarithms of half-lives for the emission of $^{12,14}$C clusters from
different deformed and superdeformed Nd and Gd parents. Our calculations, using {\it AZ-formula}
(AZF), are compared with the calculations based on PCM and the scaling law of Horoi {\it et al.}
\cite{hor94}.
\end{description}

\newpage
\begin{table*}
\caption{The logarithms of decay half-lives of different clusters
emitted from various radioactive nuclei, calculated by using the
semiempirical {\it AZ-formula} (\ref{eq:3}) and the scaling law
(\ref{eq:1}) compared with the experimental data. The experimental
Q-values are also given and the data are tabulated in the
increasing order of the mass number of the clusters. The last
column gives the reference to the source of experimental data.}
\begin{tabular}{cccccccc}\hline
Parent      &Cluster    &Daughter   &$Q^{Expt.}$ (MeV)     &$\log
T_{1/2}^{AZF}$ (s)& $\log T_{1/2}^{Ref.[15]}$(s)  &$\log
T_{1/2}^{Expt.}$ (s)&Ref. \\ \hline
$^{221}Fr$  &   $^{14}C$    &   $^{207}Tl$  &   31.28   &   14.54   &   13.56   &   14.52   &   \cite{bon94}    \\
$^{221}Ra$  &   $^{14}C$    &   $^{207}Pb$  &   32.39   &   12.96   &   12.28   &   13.39   &   \cite{bon94}    \\
$^{222}Ra$  &   $^{14}C$    &   $^{208}Pb$  &   33.05   &   11.98   &   11.00   &   11.01   &   \cite{pri85}    \\
$^{223}Ra$  &   $^{14}C$    &   $^{209}Pb$  &   31.85   &   13.81   &   13.38   &   15.20   &   \cite{pri85}    \\
$^{224}Ra$  &   $^{14}C$    &   $^{210}Pb$  &   30.54   &   15.94   &   16.13   &   15.68   &   \cite{hou91}    \\
$^{225}Ac$  &   $^{14}C$    &   $^{211}Bi$  &   30.48   &   16.20   &   17.26   &   17.16   &   \cite{bon93}    \\
$^{226}Ra$  &   $^{14}C$    &   $^{212}Pb$  &   28.21   &   20.08   &   21.50   &   21.19   &   \cite{hou85}    \\
$^{228}Th$  &   $^{20}O$    &   $^{208}Pb$  &   44.72   &   21.90   &   21.20   &   20.72   &   \cite{bon93a}   \\
$^{230}U$   &   $^{22}Ne$   &   $^{208}Pb$  &   61.59   &   21.78   &   19.28   &   20.14   &   \cite{qia00}    \\
$^{231}Pa$  &   $^{23}F$    &   $^{208}Pb$  &   51.84   &   24.30   &   23.85   &   26.02   &   \cite{pri92}    \\
$^{230}Th$  &   $^{24}Ne$   &   $^{206}Hg$  &   57.78   &   25.77   &   23.88   &   24.61   &   \cite{tre85}    \\
$^{231}Pa$  &   $^{24}Ne$   &   $^{207}Tl$  &   60.42   &   23.62   &   21.30   &   23.23   &   \cite{san84}    \\
$^{232}U$   &   $^{24}Ne$   &   $^{208}Pb$  &   62.31   &   22.24   &   19.94   &   21.08   &   \cite{bar85}    \\
$^{233}U$   &   $^{24}Ne$   &   $^{209}Pb$  &   60.5    &   23.87   &   22.53   &   24.83   &   \cite{tre85}    \\
$^{234}U$   &   $^{24}Ne$   &   $^{210}Pb$  &   58.84   &   25.42   &   25.01   &   25.92   &   \cite{bon91,moo89}  \\
$^{235}U$   &   $^{24}Ne$   &   $^{211}Pb$  &   57.36   &   26.87   &   27.31   &   27.42   &   \cite{bon91,tre89}  \\
$^{233}U$   &   $^{25}Ne$   &   $^{208}Pb$  &   60.75   &   24.15   &   22.98   &   24.83   &   \cite{tre85}    \\
$^{235}U$   &   $^{25}Ne$   &   $^{210}Pb$  &   57.83   &   26.94   &   27.49   &   27.42   &   \cite{bon91,tre89}  \\
$^{234}U$   &   $^{26}Ne$   &   $^{208}Pb$  &   59.47   &   25.85   &   25.75   &   25.92   &   \cite{bon91,moo89}  \\
$^{234}U$   &   $^{28}Mg$   &   $^{206}Hg$  &   74.13   &   26.24   &   24.74   &   27.54   &   \cite{tre89}    \\
$^{236}Pu$  &   $^{28}Mg$   &   $^{208}Pb$  &   79.67   &   23.01   &   20.83   &   21.67   &   \cite{ogl90}    \\
$^{236}U$   &   $^{28}Mg$   &   $^{208}Hg$  &   71.69   &   28.18   &   27.98   &   27.58   &   \cite{tre94}    \\
$^{238}Pu$  &   $^{28}Mg$   &   $^{210}Pb$  &   75.93   &   25.70   &   25.39   &   25.70   &   \cite{wan89}    \\
$^{236}U$   &   $^{30}Mg$   &   $^{206}Hg$  &   72.51   &   28.36   &   28.36   &   27.58   &   \cite{tre94}    \\
$^{238}Pu$  &   $^{30}Mg$   &   $^{208}Pb$  &   77.03   &   25.71   &   25.41   &   25.70   &   \cite{wan89}    \\
$^{238}Pu$  &   $^{32}Si$   &   $^{206}Hg$  &   91.21   &   25.99   &   25.68   &   25.27   &   \cite{wan89}\\  \\  \hline
\end{tabular}
\end{table*}

\end{document}